# EXPLORATION OF PARAMETERS THAT AFFECT HIGH FIELD Q-SLOPE*

K. Howard†, Y.-K. Kim, University of Chicago, Chicago, IL, USA
D. Bafia, A. Grassellino, Fermi National Accelerator Laboratory, Batavia, IL, USA

## Abstract

The onset of high field Q-slope (HFQS) around 25 MV/m prevents cavities in electropolished (EP) condition from reaching high quality factors at high gradients due to the precipitation of niobium hydrides during cooldown. These hydrides are non-superconducting at 2 K, and contribute to losses such as Q disease and HFQS. We are interested in exploring the parameters that affect the behavior of HFQS. We study a high RRR cavity that received an 800 °C by 3 hour bake and EP treatment to observe HFQS. First, we explore the effect of trapped magnetic flux. The cavity is tested after cooling slowly through $T_c$ while applying various levels of ambient field. We observe the onset of the HFQS and correlate this behavior with the amount of trapped flux. Next, we investigate the effect of the size/concentration of hydrides. The cavity is tested after holding the temperature at 100 K for 14 hours during the cooldown to promote the growth of hydrides. We can correlate the behavior of the HFQS with the increased hydride concentration. Our results will help further the understanding of the mechanism of HFQS.

## INTRODUCTION

The high field Q-slope (HFQS) observed in electropolished (EP) cavities has a typical onset around 25 MV/m. This effect is from the precipitation of niobium hydrides during cooldown and prevents EP cavities from reaching high $Q_0$'s at high gradients [1, 2]. Hydrogen is an unavoidable impurity, even in high RRR niobium. Niobium hydrides are non-superconducting at 2 K, and contribute to losses such at "Q disease" and HFQS. A 800 °C bake mitigates the Q disease, and LTB mitigates the growth of hydrides, preventing the HFQS.

In this study, we investigate a high RRR single-cell TESLA-shaped 1.3 GHz cavity. The cavity receives a 800 °C by 3 hour bake to isolate the HFQS without Q disease. Then, the cavity receives EP treatment to make the surface layer and bulk uniform [3]. Before testing, we hold the cavity at 100 K to promote the growth of hydrides. Then we perform a fast cooldown. During RF testing, we observe the behavior of the HFQS. Because the morphology of hydrides grown at these temperatures is understood [2], we can correlate the behavior of the HFQS to the hydride size/concentration.

During cavity testing, a fast cooldown is typically performed to prevent trapped magnetic flux, which is known to harm performance by increasing the residual resistance [4–9]. By not following the fast cooldown procedure, flux may be trapped through the incomplete Meissner effect, where there are normal conducting vortices within the superconducting lattice [10]. The oscillation of such normal conducting vortices in niobium during RF operation introduces significant dissipation, limiting the $Q_0$ [11, 12]. The sensitivity to trapped flux of surface treatments such as LTB and N-doping has been studied [6, 7, 9], but its effect, if any, on hydrides is not well understood. The cavity will be tested after cooling slowly through the superconducting transition while applying various levels of ambient field. Then, we observe the onset and slope of the HFQS and correlate this behavior with the amount of flux trapped.

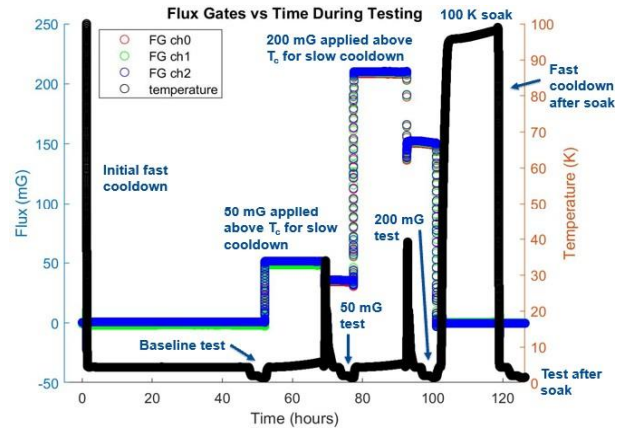

Figure 1: Measurements from flux gates and temperature sensors during testing.

The test procedure is shown in Fig. 1. In real time, the flux testing occurred before the hydride testing so that we could complete the testing in one session. We started with a fast cooldown followed by the baseline test. We measure $Q_0$ versus gradient at 2 K and low temperature (< 1.5 K) in the vertical test stand [13]. We measure the $Q_0$ at a given gradient by maintaining the cavity at its resonant frequency, inputting power via antenna, and then measuring the reflected and transmitted power [14]. The $Q_0$ is the ratio of the energy gain per RF period and dissipated power. The surface resistance is the geometry factor of the cavity divided by the $Q_0$; this can be broken down into the residual resistance ($R_{res}$) and $R_{BCS}$. The residual resistance ($R_{res}$) taken at low temperature is temperature-independent, and comes from impurities in the superconducting lattice as well as any trapped flux from cooldown or quench. The $R_{BCS}$ is calculated by taking the difference between the total surface resistance at 2 K and low T. This temperature-dependent component of the resistance is caused by the breakdown of cooper pairs with increasing temperature [15, 16]. Temperature maps are taken approximately each MV/m during the test to observe heating and quench.

* This manuscript has been authored by Fermi Research Alliance, LLC under Contract No. DE-AC02-07CH11359 with the U.S. Department of Energy, Office of Science, Office of High Energy Physics. This work was supported by the University of Chicago.
† khoward99@uchicago.edu

After baseline testing, we warm up above $T_c$ but below the hydride growth zone. We then perform a slow cooldown in a magnetic field of 50 mG. The same testing procedures are followed for the 50 mG test. We repeat this process for 200 mG. Next, we warm up to 100 K to perform a 14 hour soak to promote the growth of hydrides. A fast cooldown is then performed to test the hydride condition without trapped flux.

## RESULTS

### Effect of Size/Concentration of Hydrides

The measurements of $Q_0$ at 2 K are graphed in Fig. 2. The quality factor before and after the 100 K soak are identical. The growth of hydrides does not degrade the performance.

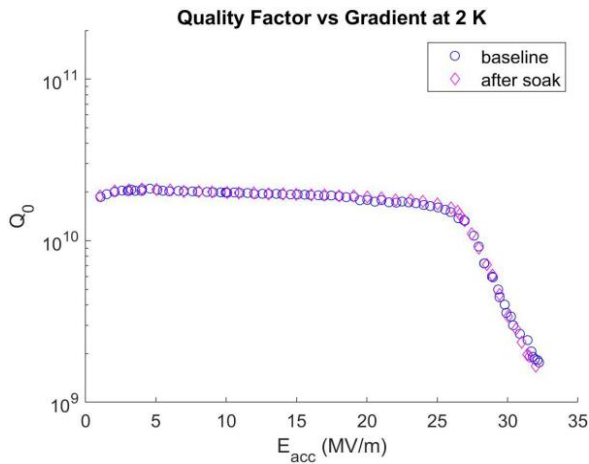

Figure 2: Quality factor at 2 K versus accelerating gradient before and after 100 K soak.

Breaking down the surface resistance into its residual and BCS components, we observe the residual resistance in Fig. 3. The residual resistance before and after the soak are identical. The BCS resistance in Fig. 4 are also identical before and after the soak. There is some weird behavior at high gradient, which may be attributed to higher uncertainties of measurement and the interpolation method used to calculate the BCS resistance. The hydride growth does not affect the RF behavior of the cavity.

Using the temperature maps taken during testing, we observed the increase in temperature with respect to the zero-field measurements. We took the average of sensors along the equator and the average of the observed hot spots which are shown in Fig. 5. There is some small variation at low fields, but the heating is nearly identical before and after the soak. The hydride growth does not affect the heating up to quench in the cavity.

Because we observe no difference before and after the 100 K soak, we believe that the 800 °C baking treatment protects against additional hydride losses. Any potential hydride growth during the soak does not affect the HFQS. The 800 °C bake allows the cavity to be robust against different cooling procedures. Because of this, the trapped flux measurements will not be affected by any hydride growth during a slow cooldown.

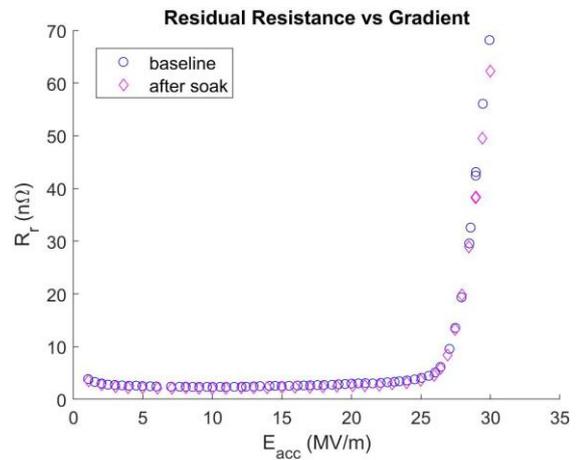

Figure 3: Residual resistance (at low T) versus accelerating gradient before and after 100 K soak.

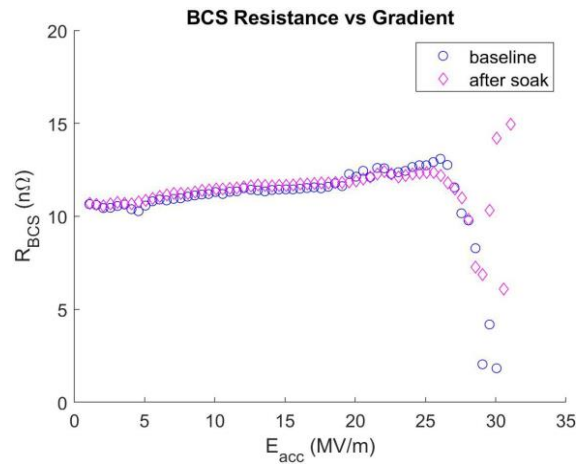

Figure 4: BCS resistance at 2 K versus accelerating gradient before and after 100 K soak.

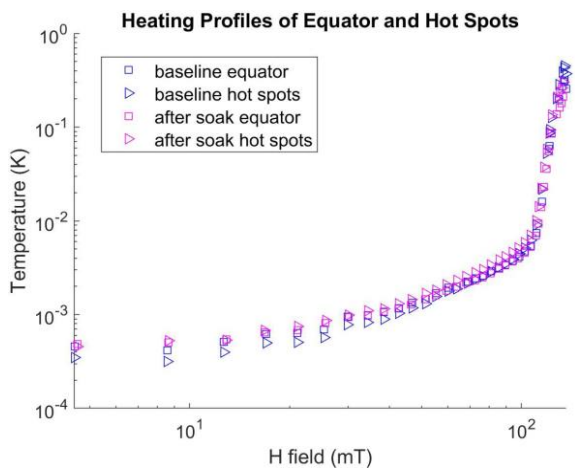

Figure 5: Heating (with respect to zero-field measurement) versus magnetic field along the equator and at hot spots before and after 100 K soak.

*Effect of Trapped Magnetic Flux*

The measurements of $Q_0$ at low temperature are graphed in Fig. 6. Focusing on the baseline measurement, we note the traditional HFQS onset at around 26 MV/m. As we increase the magnetic flux to 50 mG and then 200 mG, we see the quality factor significantly decrease. An important note is that these measurements were power-limited, not quench limited. There is no significant difference in the HFQS from what we observe.

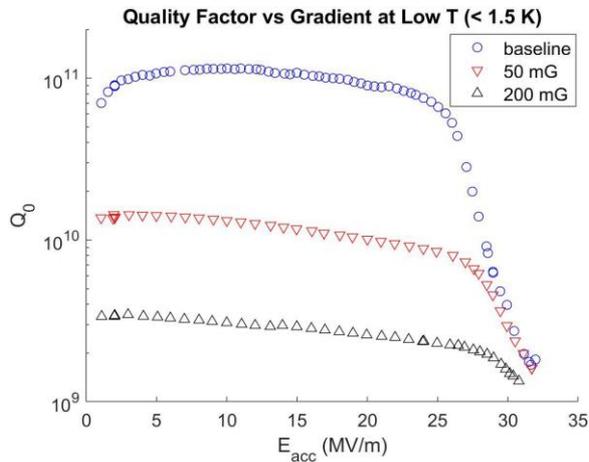

Figure 6: Quality factor at low T (< 1.5 K versus accelerating gradient at 0 mG, 50 mG, and 200 mG.

In Fig. 7, we observe the drastic increase in residual resistance from increasing the trapped magnetic flux. At high fields, the slope of the curves are similar, which is consistent with having the same HFQS behavior.

The BCS resistance, shown in Fig. 8, is the same across different levels of trapped flux. At 200 mG, we note a high uncertainty in the calculated BCS resistance because the residual resistance is such a large component of the surface resistance. At high fields, the decrease in BCS resistance is an artifact of measurement, because the uncertainty of the quality factor measurements at high fields is larger, and the difference between the quality factor at 2 K and < 1.5 K is small here. The trapped magnetic flux does not increase the BCS resistance, which is consistent with popular belief.

We observe the heating profiles averaged at the equator and the hot spots for each test in Fig. 9. At 50 mG and 200 mG, we observe that different amount of trapped flux causes the heating to be different at the equator versus the hot spots. The hot spots increase in temperature from their zero-field value much faster, whereas there is no difference in the baseline measurement. While heating before the HFQS is different with different amounts of trapped flux, we notice that all curves collapse to the same slope at the onset of he HFQS. As a result, we conclude that trapped magnetic flux does not contribute toward HFQS losses.

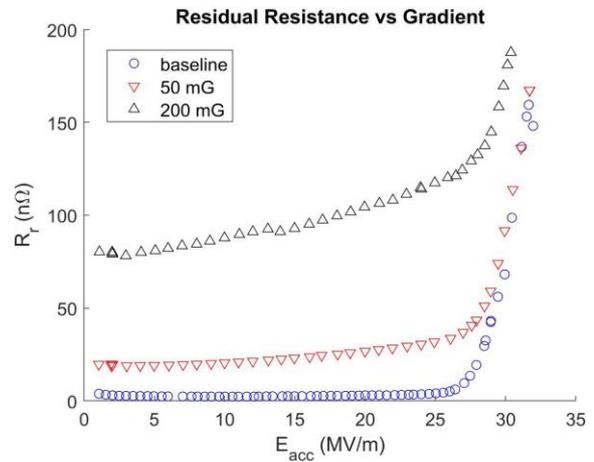

Figure 7: Residual resistance (at low T) versus accelerating gradient at 0 mG, 50 mG, and 200 mG.

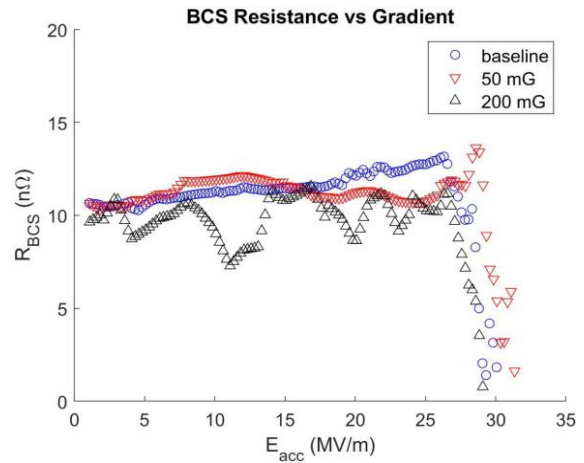

Figure 8: BCS resistance at 2 K versus accelerating gradient at 0 mG, 50 mG, and 200 mG.

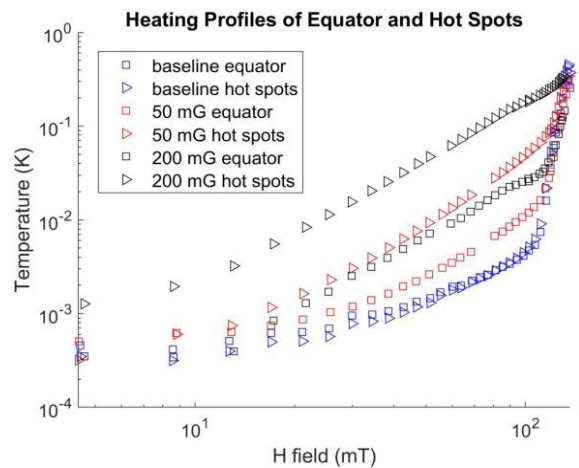

Figure 9: Heating (with respect to zero-field measurement) versus magnetic field along the equator and at hot spots at 0 mG, 50 mG, and 200 mG.

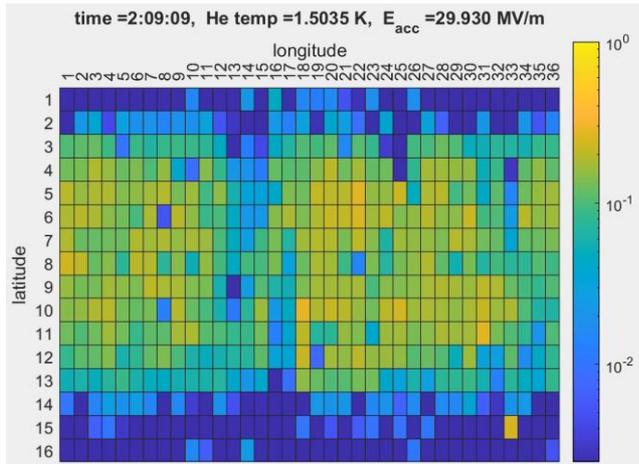

Figure 10: Temperature map of baseline test (0 mG) near 30 MV/m.

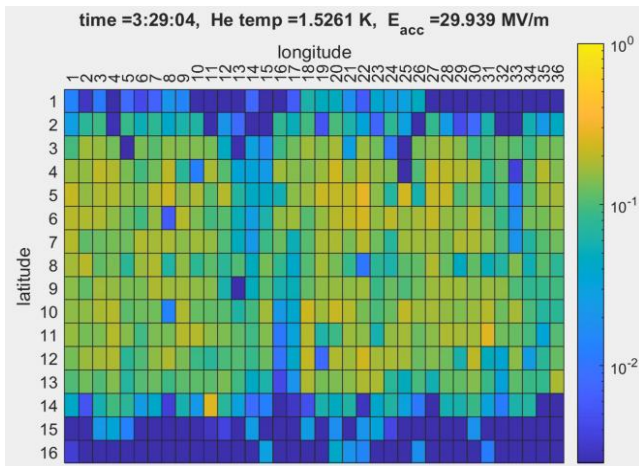

Figure 11: Temperature map of 50 mG test near 30 MV/m.

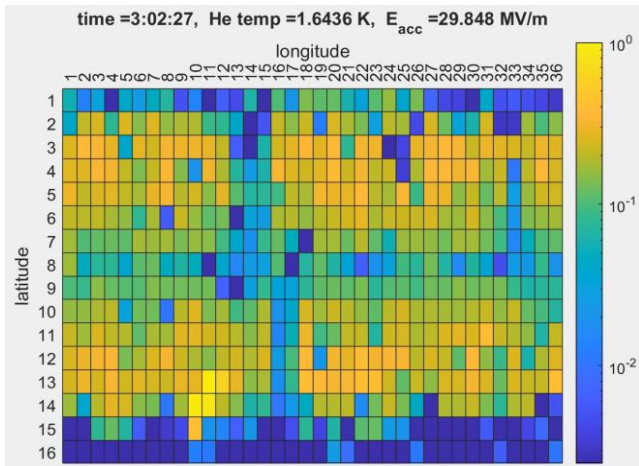

Figure 12: Temperature map of 200 mG test near 30 MV/m.

In Figs. 10, 11, and 12, we observe the temperature maps taken approximately at 30 MV/m during the low temperature portion of each test. In each map, we see the same hot spots occur, but to different intensities. The change from the baseline to 50 mG is not dramatic. However, for the 200 mG test, both the bath temperature and the hot spot intensities are significantly higher, showing the additional heating caused by the trapped magnetic flux. While there is no evidence of additional HFQS losses caused by trapped magnetic flux, the additional heating prior to the HFQS certainly affects the temperature of the cavity at high gradient.

## CONCLUSION

In this study, we explored niobium hydrides and trapped magnetic flux as potential influences on the HFQS. Through a 100 K soak that would promote the growth of hydrides in an EP cavity, we found no difference in cavity performance after 14 hours of soaking. We conclude that the 800 °C bake the cavity received in preparation for testing is robust as a protection against additional hydride-based losses. Putting the cavity in hydride growth conditions may either prevent the growth of hydrides or alter their behavior in the RF layer to mitigate losses. This is the subject of further study.

Since this cavity is unaffected by hydride growth during cooldown, we can be confident that the testing of different levels of trapped magnetic flux will be unaffected by additional hydride-based losses. This means we could isolate the effect of the HFQS. By testing the cavity in ambient magnetic fields of 50 mG and 200 mG, we observe the expected increase in residual resistance and consistency of the BCS resistance. Interestingly, we observe different heating behavior at the different levels of flux along the equator and at hot spots, but this difference collapses at the onset of the HFQS. As a result, we conclude that the levels trapped magnetic flux studied do not affect the HFQS. Limitations in power due to the high residual resistance limit our ability to observe the quench behavior at different flux levels. Because of the different levels of heating, this will be important to observe in temperature mapping in the future.